\magnification=1200
\parskip=0truept
\baselineskip=24pt 
\def\gray{{$\gamma$-ray}}
\def\grays{{$\gamma$-rays}}
\def\etal{{\it et al.}}

\def\mn{{\it Mon. Not. Royal Astron. Soc.}}
\def\apj{{\it Astrophys. J.}}
\def\apjl{{\it Astrophys. J. (Lett.)}}

\def\xxiicrc{{\it 21st Internat. Cosmic Ray Conf. Papers}}
\def\aa{{\it Astron. \& Astr.}}
\def\ref{\noindent\hangafter=1\hangindent=1truecm}
\def\mic{$\mu$m}

\centerline{\bf ON THE ABSORPTION OF HIGH ENERGY GAMMA-RAYS}
\centerline{\bf BY INTERGALACTIC INFRARED RADIATION}
\vskip 18pt
\centerline{F. W. STECKER}
\centerline{Laboratory for High Energy Astrophysics}
\centerline{NASA Goddard Space Flight Center,Greenbelt, MD 20771, U.S.A.}
\vskip 12pt
\centerline{AND}
\vskip 12pt
\centerline{O.C. DE JAGER}
\centerline{Department of Physics, Potchefstroom University}
\centerline{Potchefstroom 2520, South Africa}

\vfill\eject

\centerline{ABSTRACT}

We present a new calculation of the intergalactic \gray\ pair-production 
absorption coefficient as a function of both energy and redshift up to the
redshift of 3C279, z = 0.54. In reexamining this problem, we
make use of new observational data on the intergalactic infrared 
radiation field (IIRF), together with recent theoretical models of 
the galactic spectral energy distributions of the IIRF from stars and
dust reradiation and estimates of the IIRF from galaxy counts and {\it COBE}
results. We present
our results for two fairly well defined IIRF spectral energy distributions, 
one of which is within $1 \sigma $ of our previous estimate 
of the IIRF at $ \sim 20$ \mic.
We then apply our results to the \gray\ spectrum of Mrk 421, and obtain good
agreement with the observational data, including the recent results of the
{\it HEGRA} group.
\vskip 18pt
\noindent{{\it Subject headings:} \grays:theory -- infrared:general --
quasars:general -- quasars:individual (Markarian 421, 3C279)
\vfill\eject

\noindent{\bf 1. Introduction}
\vskip 14pt
The magnitude and shape of the intergalactic infrared radiation field
(IIRF) are fundamental to understanding the 
early evolution of galaxies. 
The IIRF is determined by star formation processes
occurring very early in the evolution of galaxies - emission of radiation
from the photospheres of cool stars, dust reradiation, and redshifting to 
even longer wavelengths. The IIRF is especially sensitive to rates of early
star formation, since much of the energy in starbursts escapes in the infrared.
The study of the IIRF is therefore of great cosmological significance.

We have previously pointed out (Stecker, De Jager and Salamon 1992 (Paper I)) 
that very high energy \gray\ beams from blazars can be used to measure  the
intergalactic infrared radiation field, since 
pair-production interactions of \grays\ with intergalactic IR photons 
will attenuate the high-energy ends of blazar 
spectra. Determining the intergalactic IR field, in turn, allows us to 
model the evolution of the galaxies which produce it. 
As energy thresholds are lowered 
in both existing and planned ground-based
air Cherenkov light detectors (Cresti 1996), cutoffs in the \gray\ spectra of 
more distant blazars are expected, owing to extinction by the IIRF. These
can be used to explore the redshift dependence of the IIRF. 
Furthermore, by using blazars for a determination 
of attenuation as a function of
redshift,  combined with a direct observation of the IR background from
the {\it DIRBE}  detector on {\it COBE}, one can, in principle, 
measure of the Hubble
constant $H_{0}$ at truly cosmological distances (Salamon, Stecker and
De Jager 1994). 

The potential importance of the photon-photon pair-production process in 
high energy astrophysics was first pointed out by Nikishov (1961). 
Unfortunately, his early paper overestimated the energy density of the  
IIRF by about three orders of magnitude. However, with the discovery 
of the cosmic microwave background radiation, Jelley (1966) and 
Gould and Schreder (1967) were quick to point out the opacity 
of the universe to photons 
of energy greater than 100 TeV. Stecker (1969) and Fazio and Stecker (1970) 
generalized these calculations to high redshifts, showing that photons 
originating at a redshift $z$ will be absorbed above an energy of $ \sim 
100 (1+z)^{-2}$ TeV. 

An {\it EGRET}--strength ``grazar'' (observed \gray\
blazar) with a hard spectrum extending to multi-TeV energies is potentially
detectable with ground-based telescopes (see the review on the
Atmospheric Cherenkov Technique by 
Weekes (1988)). Pair-production interactions with
the cosmic microwave background radiation (CBR) will not cut off the \gray\
spectrum of a low-redshift source below an energy of $\sim 100$
TeV (see above). However, as pointed out in Paper I, even 
bright blazars at moderate redshifts ($z >0.1$) will suffer absorption at TeV
energies owing to interactions with the IIRF.

Shortly after Paper I was published, the Whipple Observatory group reported 
the discovery of $\sim$ TeV \grays\ coming from the blazar Mrk 421, a
BL Lac object with a redshift of 0.031 (Punch \etal\ 1992). This 
source has a hard, roughly $E^{-2}$,
\gray\ spectrum extending to the TeV energy range (Mohanty, \etal\ 1993; 
Lin, \etal\ 1994) 
where the pair-production process involving infrared
photons becomes relevant. Following this discovery, we made use
of the TeV observations of Mrk 421 to carry out our suggestion of using
such sources to probe the IIRF (Stecker and De Jager 1993 (Paper II);
De Jager, Stecker and Salamon 1994). This problem was also discussed by Dwek and Slavin (1994) and Biller, \etal\ (1995). 

There are now over 50 grazars which have been detected by the {\it EGRET} team
(Thompson, \etal\ 1995). These sources, optically violent variable quasars
and BL Lac objects, have been detected out to a redshift greater that 2.

Of all of the blazars detected by {\it EGRET}, only the low-redshift 
BL Lac, Mrk 421, has been seen by
the Whipple telescope. The fact that the Whipple team did not detect the
much brighter {\it EGRET} source, 3C279, at TeV energies (Vacanti, \etal\ 1990,
Kerrick, \etal\ 1993) is consistent with the predictions of a
cutoff for a source at its much higher redshift of 0.54 (see Paper I).
So too is the recent observation of two other very close BL Lacs ($z < 0.05$),
{\it viz.}, Mrk 501 (Quinn, \etal\ 1996) and 1ES2344+514 (Schubnell 1996)
which were too faint at GeV energies to be seen by {\it EGRET}.

In this paper, we calculate the absoption coefficient of intergalactic
space using recent theoretical
models of the spectral energy distribution (SED) of intergalactic low energy 
photons in conjunction with recent 
determinations of the
IR background obtained by analysis of {\it COBE} data and galaxy counts.
After giving our results on the \gray\ optical depth as a function of energy 
and redshift out to a redshift of 0.54, we apply our calculations 
to the high-energy \gray\ spectrum of Mrk 421 and 
compare our results with the published spectral data obtained by 
the {\it HEGRA} group (Petry, \etal\ 1996) and the preliminary spectrum
obtained by the Whipple group (Mohanty,\etal\ 1993).

\vfill\eject
\noindent{\bf 2. Absorption of High Energy Gamma-Rays by Low Energy Photons  
in Intergalactic Space}

\vskip 14pt

The formulae relevant to calculations involving the pair-production process 
are given and discussed in Paper I.
For $\gamma$-rays in the TeV energy range, the pair-production cross section 
is maximized
when the soft photon energy is in the infrared range: 
$$\lambda (E_{\gamma}) \simeq \lambda_{e}{E_{\gamma}\over{2m_{e}c^{2}}} =
2.4E_{\gamma,TeV} \; \; \mu m \eqno{(1)}$$ where $\lambda_{e} = h/(m_{e}c)$ 
is the Compton wavelength of the electron.
For a 1 TeV $\gamma$-ray, this corresponds to a soft photon having a
wavelength  near the K-band (2.2\mic). (Pair-production interactions actually
take place with photons over a range of wavelengths around the optimal value as
determined by the energy dependence of the cross section.) 
If the emission spectrum of
an extragalactic source extends beyond 1 TeV, then the extragalactic
infrared field should cut off the {\it observed} spectrum between $\sim
10$ GeV and a few TeV, depending on the redshift of the source.  

The TeV spectrum of Mrk 421 obtained by the Whipple Observatory group is
consistent with an $E^{-2}$ power-law, showing no significant
absorption out to an energy of at least 3 TeV (Lin \etal\ 1994; 
Sreekumar, \etal\ 1996; Mohanty \etal\ 1993).  It is
this lack of absorption below 3 TeV which we (Stecker and De Jager 1993)
used to put upper limits on the extragalactic infrared energy density in
the $\sim$ 1 $\mu$m to 8 $\mu$m wavelength range (see
also Dwek and Slavin 1994). 

Our upper limits on the IIRF, obtained from the Whipple data
on Mrk 421 below 3 TeV, ruled out various exotic mechanisms for producing
larger fluxes, such as decaying particles, exploding stars, massive
object and black hole models (Carr 1988). They are consistent with the
extragalactic near infrared background originating from ordinary stellar
processes in galaxies (Stecker, Puget and Fazio 1977; 
Franceschini \etal\ 1994; see also the papers included in Dwek 1996). 

As pointed out by Dwek and Slavin (1994), the 
SED of the IIRF should consist of both starlight and dust reradiation 
components with the later component peaking in the far infrared region.
Thus, the power-law SEDs which we used previously are approximations
which are valid over a limited wavelength range.
In this paper we utilize the SED models of Franceschini \etal\ 1994, 
together with
recent data, in order to make a better determination of the IR 
background in intergalactic space and to calculate the \gray\ absorption
coefficient from pair-production off the IIRF. 
  
\vskip 18pt
\noindent{\bf 3. The Intergalactic Infrared Radiation Field}
\vskip 14pt
A theoretical upper limit to the intergalactic infrared energy density 
whose ultimate origin is stellar nucleosynthesis was given by Stecker \etal\
(1977). Direct observational limits on the IIRF, determined by
analysis of the {\it COBE/DIRBE} data with other components 
subtracted out, has been
given by Hauser (1996). Both of these determinations are well below the limits
obtained by Biller, \etal\ (1995). Biller, \etal\ (1995) 
argued that the most conservative estimate
of an upper limit on the IIRF would be obtained by taking the flattest
spectral index for Mrk 421 consistent with {\it EGRET} data at GeV energies
and assuming that the spectrum of Mohanty et al. (1993) is the result
of absorption with an optical depth $\tau > 6$ in the TeV energy range.
The huge IIRF required to obtain this optical depth is inconsistent with the 
{\it COBE/DIRBE} data (Hauser 1996). Also, in order to
produce a power-law absorbed \gray\ spectrum from a different power-law 
source spectrum in the TeV energy range, they had
to employ an unphysical SED which is inconsistent with that produced by
theoretical models of stellar and dust emission from galaxies.

Recently, Puget, \etal\ (1996) have claimed a tentative detection of
the far-infrared background from an analysis of {\it COBE} data at wavelengths
greater than 200 \mic. An estimate of the IIRF from counts of {\it IRAS} 
galaxies has been given by Gregorich, \etal\ (1995). We have given both
estimates of and upper limits to the IIRF (Stecker and De Jager 1993;
De Jager, Stecker and Salamon 1994) based on analyses of the preliminary
spectrum of Mrk 421 at TeV energies obtained by the Whipple group (Mohanty,
\etal\ 1993).
All of these direct and indirect observational 
results can be brought together into a coherent
scenario where the IIRF is assumed to be generated by stellar processes in
galaxies. The resulting IIRF from both starlight and dust reradiation should
then have a two-peaked SED made up of the contributions from all of the
galaxies in the universe. The energy flux of the IIRF is
expected to be within a factor of two of the value of 10 nWm$^{-2}$sr$^{-1}$.
In Figure 1, we have plotted the observational data together with two 
theoretical SEDs of the galaxy-generated IIRF. 
These SEDs, the thick solid lines 1 and 2, are based on the models of 
Franceschini, \etal\ (1994). We used the starlight component exactly as
given by these authors; However, the more uncertain dust component was
normalized to a slightly higher value by
using the data of Puget, \etal\ (1996) in the far-infrared range for their
two different assumptions about subtraction of HII emission. 
The estimate of the IIRF from {\it IRAS} galaxy counts by Grigorich, \etal\
(1995) at 60 \mic\ is consistent with both models within its uncertainty.

We also show
an estimate of the SED at intermediate wavelengths from dust tori around
Seyfert galaxy nuclei by Granato, \etal\ (1996). This additional component
could add significantly to the lower SED model (model 1) at 
wavelengths between 10 \mic\ and 100 \mic. 

Our chosen SEDs are generally within a factor of 2 of each
other over the wavelength range of significance for the absorption 
of TeV \grays.
We therefore feel that they represent, within the narrowest range of error
presently available, a reasonable construction of the IIRF. We have used
these two SEDs to calculate the opacity of intergalactic space to high 
energy \grays\ and to apply these calculations to observations of Mrk 421
in particular.

\vskip 18pt
\noindent{\bf 4. The Opacity of Intergalactic Space to the IIRF}
\vskip 14pt

The technique for calculating the pair-production optical 
depth as a function of energy has been given in Papers I and II. We
do not here take account of changes in the IIRF with redshift, and
instead make the assumption that the background is basically in
place at a redshifts $<$ 0.54, having been produced primarily at higher
redshifts (Cowie, \etal\ 1996). 
We therefore limited our calculations to $z<0.54$, the redshift of the
powerful grazar 3C279 which we discussed in Paper I.

Figure 2 shows the results of our calculations of the optical depth for 
various energies and redshifts up to 0.54, the redshift of the powerful
{\it EGRET} grazar 3C279.
We take a Hubble constant of $H_o=70$ km s$^{-1}$Mpc$^{-1}$,
which is consistent with various recent observations as presented at the
recent Space Telescope Symposium on the Extragalactic Distance Scale.
We assume for the IIRF, the two model SEDs given by the
solid lines marked 1 (lower curve) and 2 (upper curve) in Fig. 1. 
The two model SEDs yield different values for the optical depth 
$\tau (E)$ only for $E>1$ TeV, since the SEDs only differ at
wavelengths greater than 2 \mic. We note that $\tau$ is greater than
1 for 3C279 for energies above $\sim$100 GeV. Designing air
Cherenkov telescopes with lower threshold energies should enable these
sources to be detectable.

We have obtained parametric expressions for  $\tau_1(E)$ and
$\tau_2(E)$ (corresponding to models 1 and 2 respectively) for
$z<0.3$. The double-peaked form of the SED of the IIRF requires
a 4th order polynomial to reproduce parametrically. It is of the form
$$log_{10}[\tau_j(E_{\rm TeV},z)]=\sum_{i=0}^4(a_{ij}+b_{ij}\log_{10}{z})
(\log_{10}E_{\rm TeV})^i\;\;{\rm for}\;\;0.1<E_{\rm TeV}<50 \eqno{(2)}$$
for models $j=1$ and 2. Table 1 gives the best-fit coefficients
$a_{ij}$ and $b_{ij}$ for each model.
These parametric expressions can be used by observers 
to correct for intergalactic absorption
and derive the true spectrum of sources out to a redshift of 0.3. 

Figure 3 shows spectra from various instruments during the quiescent
phase of Mrk 421 compared with the spectra calculated including absorption 
predictions for the two SEDs which we used for the IIRF.
The {\it EGRET} differential spectrum of Mrk 421 (Sreekumar, \etal\ 1996)
was converted to an integral spectrum in the range 70 MeV to 4 GeV,
after adding the integral flux above 4 GeV to the {\it EGRET} differential
points, assuming a model spectrum given by (for $j=1$ and 2)
$$I(E_{\gamma})=1.3\times 10^{-11} 
E_{\rm TeV}^{-2}\exp{(-\tau_j(E;z=0.031))}\;\;{\rm cm^{-2}s^{-1}TeV^{-1}}.\eqno{(3)}$$
Absorption does not influence the GeV part of the spectrum.
The integral flux above 0.1 GeV implied by eq. (3) is within 1 $\sigma$
of the published flux for Mrk 421 (Sreekumar, \etal\ 1996). By forcing this
$E^{-1}$ integral spectrum through the {\it EGRET} data, we obtain a value for
$\chi^{2}$ of 7.5 for 5 degrees of freedom, showing consistency with our 
assumed spectrum. Normalizing our spectrum to the {\it EGRET} data in this
way, we find excellent agreement with the TeV data as shown in Figure 3, even
though the TeV spectra are from different time periods.

Above 1 TeV, Petry, \etal\ have 
found a steep integral spectral index of $2.8\pm 0.6$ for Mrk 421.
The preliminary Whipple spectrum for Mrk 421 
(Mohanty, \etal\ 1993) appears to be consistent with this steep
index as shown in Figure 3, even though this spectrum is highly uncertain
above 5 TeV (R.C. Lamb, private communication; Biller, \etal\ 1995).
The spectrum given by eq. (3) connects the {\it EGRET} 
data at GeV energies to the TeV spectrum of the
{\it HEGRA} collaboration (Petry, \etal\ 1996) and follows the spectral
steepening in the data above 1 TeV. 

The agreement between observation and calculation here supports our
assumption that electron-photon cascading (Protheroe and Stanev 1993;
Aharonian, Coppi and V\"{o}lk 1993) is not expected to significantly
influence the TeV spectrum of Mrk 421. This is because the source is
observed with a point-spread function of HWHM of $\sim 0.2^{\circ}$
and it can be shown that even an intergalactic magnetic field as
small as $10^{-20}$ G would deflect the cascade electrons away from
the source direction, creating a \gray\ halo much larger than the
point-spread function. Since the measured magnetic field strengths in
galaxy clusters and superclusters are of the order of a $\mu$G
(Kronberg 1994 and refs. therein), it is highly unlikely that the
intergalactic magnetic field would be as weak as $10^{-20}$ G.

\vfill\eject

\noindent{\bf 5. Conclusions}
\vskip 14pt

We have calculated the absorption coefficient of intergalactic space
from interactions with low energy photons of the IIRF
both as a function of energy and redshift, using reasonable
theoretical SEDs which are consistent with present observational
data. We have further applied our calculations to the grazar
Mrk 421, which is the only extragalactic source for
which spectral data presently exist at TeV energies.

Observations of the \gray\ spectrum of the Crab Nebula from the same 
groups which have observed Mrk 421 both show agreement that the 
integral spectral index of the Crab Nebula is $\sim$1.7 at
$\sim$ 1 TeV
(Weekes, \etal\ 1994; Petry, \etal\ 1996). Furthermore, Petry, \etal\ (1996)
have shown that the spectrum of the Crab Nebula above 1 TeV is flatter than the
corresponding spectrum of Mrk 421. 
The index of the integral spectrum of Mrk 421 of 2.8$\pm$0.6 derived
by Petry, \etal\ (1996) is significantly steeper than the value of $\sim$ 1
needed to connect the GeV and TeV data (see Figure 3). While preliminary,
the data of Mohanty, \etal\ (1993) are also consistent with this conclusion.

Thus it appears that there
is substantial observational evidence for a real steeping in the spectrum
of Mrk 421 at a several TeV (see Figure 3) which may be naturally 
interpreted as an absorption cutoff caused by pair production
interactions with the IIRF
(De Jager, Stecker and Salamon 1994). For the two SEDs which we have 
chosen for the IIRF, the lower model (model 1) would allow 
for some additional {\it intrinsic} absorption in the source. 
However the use of either model will
account for the spectral data (within the errors) without 
postulating any significant
intrinsic absorption or a cutoff in the emission spectrum of the
source at that particular energy. Both model SEDs give a value of $\tau > 1$
at \gray\ energies above 5 TeV (see Figure 2). 
Therefore, we feel that absorption
in intergalactic space by the IIRF is the simplest, most natural
explanation for the TeV spectrum of Mrk 421. It is an explanation
which is consistent with present observations of the IIRF and
the predicted absorption effect obtained using reasonable theoretical models
of the IIRF. This, in turn,
gives us confidence in our predictions of the absorption coefficient
for sources at other redshifts.

Our calculations confirm our previous conclusions in Paper I
that the high energy (TeV) spectra of sources 
at redshifts higher than 0.1 should suffer significant absorption. The recent
detections of two additional sources at redshifts below 0.1, {\it
viz.}, Mrk 501 (Quinn, \etal\ 1996) and 1ES2344+514 (Schnubell 1996),
coupled with the non-observations of {\it all} of the other known {\it
EGRET} blazars at redshifts above 0.1, even those which are much
brighter than Mrk 421 in the GeV range and have comparable spectral
indeces, further supports our calculations and interpretation of the
Mrk 421 spectrum.

\vskip 12pt
\noindent{ACKNOWLEDGEMENT:}

We wish to thank Dirk Petry for sending us his results prior to publication 
and for helpful discussions.

\vfill\eject

\noindent {\bf REFERENCES}
\vskip 18pt

\ref Aharonian, F.A., Coppi, P. and V\"{o}lk, H.J. 1993, {\it Proc. 23rd Intl. Cosmic Ray Conf.}, Univ. Calgary Press: Calgary, {\bf 1}, 451.

\ref Bertsch, D. L. \etal\ 1993, \apj\ {\it Letters} {\bf 405}, L21. 

\ref Biller, S.D., \etal\ 1995, \apj\ {\bf 445}, 227.

\ref Carr, B.J. 1988, in A. Lawrence (ed.),
{\it Comets to Cosmology}, Springer-Verlag: Berlin, p. 265.

\ref Cowie, L.L., Gardner, J.P., Lilly, S.J. and McLean, I. 1990,
\apjl {\bf 360}, L1. 

\ref Cowie, L.L. \etal\ 1996, \apj , in press.

\ref Cresti, M. ed. {\it Towards a Major Atmospheric Cherenkov Detector IV},
Padova (1996).


\ref De Jager, O.C., Stecker, F.W. and Salamon, M.H. 1994, 
{\it Nature}\ {\bf 369}, 294. 

\ref Dwek, E. 1996, (ed.) {\it Unveiling the Cosmic Infrared Background}, 
AIP CP 348, Amer. Inst. Phys.:New York. 

\ref Dwek, E. and Slavin, J. 1994, \apj, {\bf 436}, 696.


\ref Fazio, G.G. and Stecker, F.W. 1970, {\it Nature}\ {\bf 226}, 135.

\ref Franceschini, \etal\ 1994, \apj\ {\bf 427}, 140.

\ref Gould, R. J. and Schreder, G.P. 1966, 
{\it Phys. Rev. Letters}\ {\bf 16}, 252.

\ref Granato, G.L., Franceschini, A., and Danese, L. 1996, in
{\it Unveiling the Cosmic Infrared Background}, AIP CP 348, ed.
E. Dwek, Amer. Inst. Phys.:New York, 226. 

\ref Gregorich, D. T., \etal\ 1995, {\it Astron. J.}, {\bf 110}, 259.

\ref Hauser, M. G. 1996, in 
{\it Unveiling the Cosmic Infrared Background}, AIP CP 348, ed.
E. Dwek, Amer. Inst. Phys.:New York, 11. 

\ref Jelley, J.V. 1966, {\it Phys. Rev. Lett.} {\bf 16}, 479.

\ref Karle, \etal\ 1995, {\it Astropart. Phys.} {\bf 4}, 1.

\ref Kerrick, A.D. \etal\ 1993, 
{\it Proc. 23rd Int'l. Cosmic Ray Conf.},
Univ. of Calgary Press, Calgary, {\bf 1}, 405.

\ref Kronberg, P.P. 1994, {\it Rpts. Prog. Phys.}, {\bf 57}, 325.

\ref Lin, Y.C., \etal\ 1994, in {\it The 2nd Compton Symposium}, 
ed. C. Fichtel, N. Gehrels and J.P. Norris, AIP CP 304, 
Amer. Inst. Phys.:New York, 582.

\ref Mohanty, G. \etal\  1993, 
{\it Proc. 23rd Int'l. Cosmic Ray Conf.},
Univ. of Calgary Press, Calgary, {\bf 1}, 440.

\ref Nikishov, A.I. 1962, {\it Sov. Phys. JETP}\ {\bf 14}, 393.

\ref Petry, \etal\ 1996, \aa , in press.

\ref Protheroe, R.J. and Stanev, T. 1993, \mn {\bf 264}, 191.

\ref Puget, J.-L. \etal\ 1996, {\it Astron. and Ap.} {\bf 308}, L5.

\ref Punch, M. \etal\ 1992, {\it Nature}\ {\bf 358}, 477.

\ref Quinn, \etal 1996, \apj, {\bf 456}, L83.

\ref Salamon, M.H., Stecker, F.W., and de Jager, O.C., \apjl\ {\bf 423},
L1 (1994).

\ref Schubnell, M.S. 1996, invited talk at 
{\it Amer. Astr. Soc. High Energy Astrophysics
Div. Conf}, San Diego, CA.

\ref Sreekumar, P. \etal\ 1996, \apj\ {\bf 464}, 628.

\ref Stecker, F.W. and De Jager, O.C. 1993, \apjl\ {\bf 415}, L71 (Paper II).

\ref Stecker, F.W., De Jager, O.C. and Salamon, M.H. 1992,
\apjl {\bf 390}, L49 (Paper I). 

\ref Stecker, F.W., Puget, J.L. and Fazio, G.G. 1977, \apjl\
{\bf 214}, L51. 

\ref Thompson, D. J., \etal\ 1995, \apj\ {\it Suppl.}, {\bf 101}, 259.

\ref Tyson, J.A. 1990, in S. Bowyer and C. Leniert (eds.),
{\it The Galactic and Extragalactic Background
Radiation}, IAU: the Netherlands, p.245.


\ref Vacanti, G., \etal\ 1990, \xxiicrc , {\bf 2}, 329. 

\ref Weekes, T.C. 1988, {\it Physics Reports}\ {\bf 160}, 1.

\ref Weekes, T.C. \etal\ 1994, in {\it Proc. Second Compton Symp.} ed. C.E.
Fichtel, N. Gehrels, and J.P. Norris, Amer. Inst. Phys, New York, p. 270.


\vfill\eject
\centerline{Table 1}
\centerline{Coefficients for the polynomial in Eqn. (2)}
$$\vbox{\tabskip = 30pt\halign{%
\hfil#\hfil&\hfil#\hfil&\hfil#\hfil&\hfil#\hfil&\hfil#\hfil\cr
\noalign{\smallskip\hrule\smallskip}
$i$&$~a_{i1}$&$~b_{i1}$&$~a_{i2}$&$~b_{i2}$\cr
\noalign{\smallskip\hrule\smallskip}
0&~1.24& ~1.08& ~1.31& ~1.12\cr
1&~0.66& ~0.00& ~0.76& ~0.02\cr
2&-0.19& ~0.21& ~0.00& ~0.26\cr
3&~0.45&-0.27& ~0.44&-0.27\cr
4&-0.15& ~0.09&-0.19& ~0.07\cr
\noalign{\smallskip\hrule\smallskip}
}}$$

\vfill\eject

\centerline{\bf Figures}
\vskip 24pt
\noindent{\bf Fig.1} 
\vskip 12pt
The low energy intergalactic photon spectrum from the 
microwave to the optical, including the
cosmic microwave background (CMB) spectrum and dust and starlight spectra. 
The theoretical SEDs given by the thick solid lines labled 1 and 2 are
based on models of Franceschini, \etal\ (1994), slightly renormalized to the
results of Puget, \etal\ 1996 (filled circles). The filled square is the
result of Grigorich, \etal\ (1995). The upper-limit arrows are the {\it
COBE/DIRBE} residuals given by Hauser (1996), with the arrow lengths
representing the range of the residuals. The lower-limit arrow near 2 \mic\
is from Tyson (1990), based on galaxy counts. We also show our previous
upper limit (Stecker and De Jager 1993 - SD93) and estimate (De Jager,
\etal\ 1994 - DSS94) based on the Mrk 421 data from Mohanty (1993).
The upper scale shows the characteristic energy for \gray\ absorption
from photons of wavelngth $\lambda$ as given by eq. (1).
\vskip 18pt
\noindent{\bf Fig.2}
\vskip 12pt
Optical depth versus energy for \grays\ originating at various redshifts
obtained using the SED models 1 and 2 as shown in Fig. 1.
\vskip 18pt

\noindent{\bf Fig.3}
\vskip 12pt
An $E^{-2}$ power-law Mkr 421 spectrum with absorption calculated using
SED models 1 and 2 of Figure 1 and
fitted to the observational data from {\it EGRET} as described in the text. 
The Whipple data are from Mohanty, \etal\ (1993). The {\it EGRET} data are
from Sreekumar, \etal\ (1996) and Sreekumar, private communication.
The {\it HEGRA} data are from Petry, \etal\ (1996). 
The {\it AIROBICC} upper limit is from (Karle, \etal\ 1995), labled ``A''.


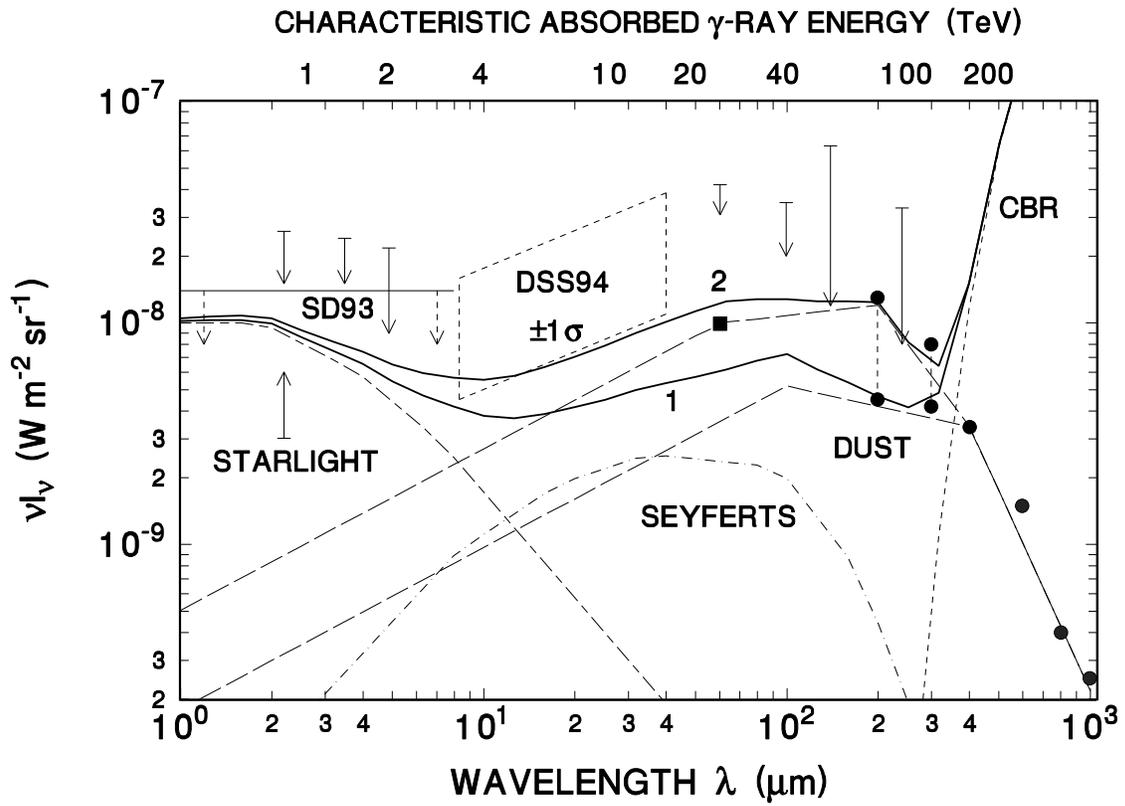
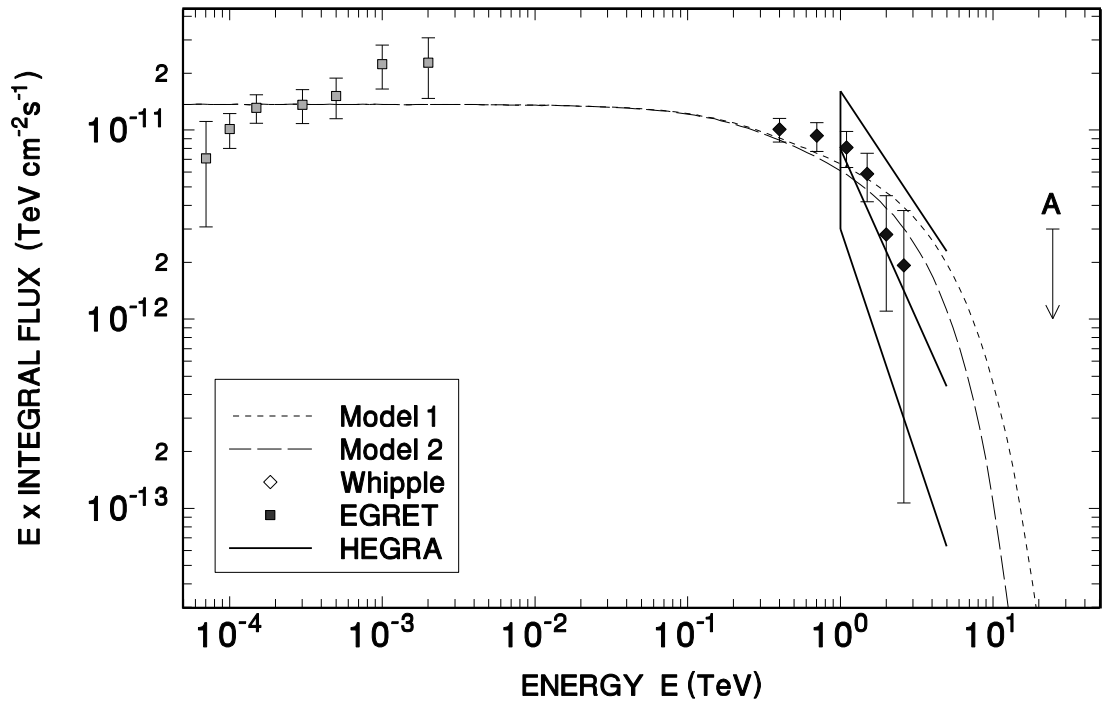
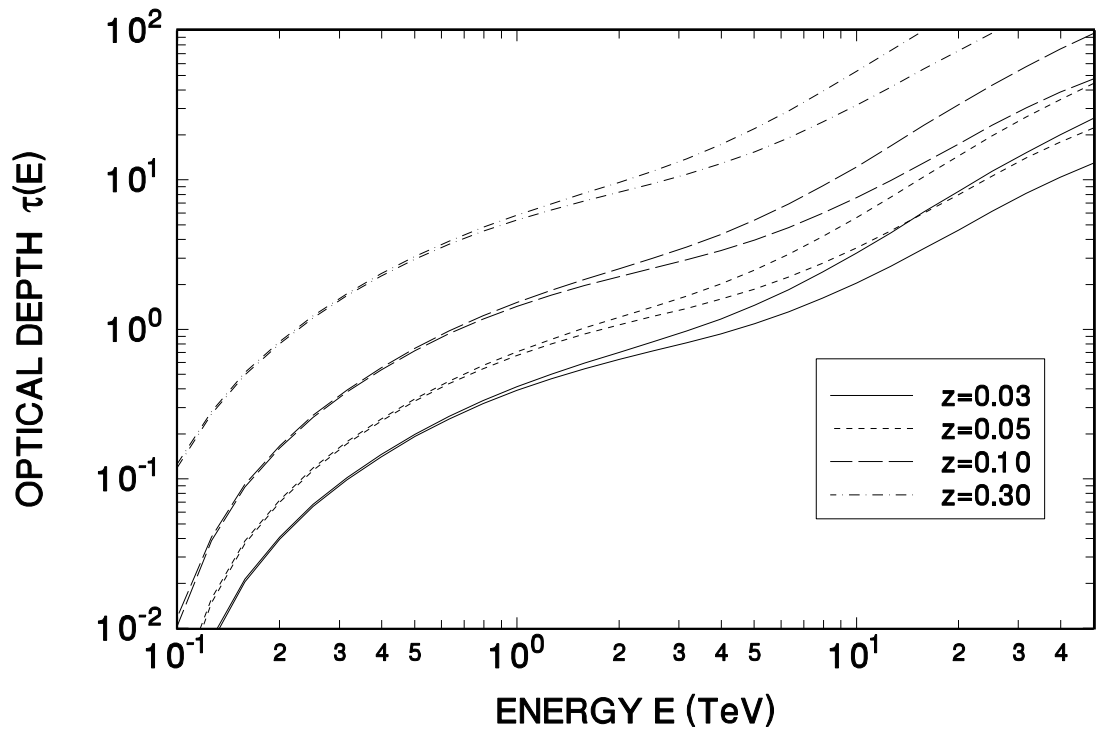
\bye